\documentclass[twocolumn,floats,floatfix,showpacs,superscriptaddress,balancelastpage]{revtex4-1}

\usepackage{graphicx}
\usepackage{amsmath}
\usepackage{amsfonts}
\usepackage{mathrsfs}
\usepackage[usenames]{color}
\usepackage[utf8]{inputenc}

\renewcommand{\vec}[1]{\mathbf{#1}}

\begin{document}

\title{Nematic braids: topological invariants and rewiring of
  disclinations}

\author{Simon \v{C}opar}

\affiliation{Faculty of Mathematics and Physics, University of
  Ljubljana, Jadranska 19, 1000 Ljubljana, Slovenia}

\author{Slobodan \v{Z}umer}

\affiliation{Faculty of Mathematics and Physics, University of
  Ljubljana, Jadranska 19, 1000 Ljubljana, Slovenia}

\affiliation{Jo\v{z}ef Stefan Institute, Jamova 39, 1000 Ljubljana,
  Slovenia}

\date{\today}


\begin{abstract}
The conventional topological description given by the fundamental
group of nematic order parameter does not adequately explain the
entangled defect line structures that have been observed in nematic
colloids.  We introduce a new topological invariant, the self-linking
number, that enables a complete classification of entangled defect
line structures in general nematics, even without particles, and
demonstrate our formalism using colloidal dimers, for which entangled
structures have been previously observed. We also unveil a simple
rewiring scheme for the orthogonal crossing of two $-1/2$
disclinations, based on a tetrahedral rotation of two relevant
disclination segments, that allows us to predict possible nematic
braids and calculate their self-linking numbers.
\end{abstract}

\maketitle

Depending on the temperature and their molecular properties,
nematogenic media can form isotropic, nematic, chiral nematic or even
blue phases, each characterized by a specific orientational ordering
of the constituent molecules. Their anisotropic nature allows the
formation of disclinations that can be stabilized by geometric or
intrinsic constraints. Recently, a lot of progress has been made on
the stabilization and manipulation of disclinations by using
dispersions of colloidal particles
\cite{poulin,musevic,lapointe,cholesteric} or confinement to porous
networks \cite{tanaka2}. In nematic dispersions, the anisotropic
inter-particle interactions mediated by elastic deformations and
defects lead to diverse colloidal structures that promote
self-assembly and offer great potential for photonics and plasmonics
\cite{phot}. Nematic braids are disclination networks where defect
loops are not localized around one particle, but instead entangle
clusters of particles \cite{zum1,zum3}. The existence of entangled
structures was first proposed based on the results of numerical
simulations \cite{keystone,araki}. They have since been observed
experimentally and their stability has been extensively analyzed
\cite{zum1,zum2}.  These structures are not sufficiently well
described by the theory developed for simple nematic defects
\cite{mermin} and a complete theoretical understanding is still
lacking. Nematic braids may also include knots and links \cite{igor},
otherwise seen in the physics of polymers \cite{polymers}, DNA
\cite{kamwrith,full,dna} and knotted light \cite{light}. Easy
experimental observation of nematic disclination networks, and their
rewiring, knotting and linking by laser tweezers \cite{zum1,igor},
places nematic braids as a primary template for the study of
nontrivial topology in physical systems.

Nematic braids stabilized by homeotropic particles consist of closed
$-1/2$ disclination loops. To fully describe a single disclination
loop, we generalize the mathematical notion of a loop by introducing
the self-linking number, which counts how many times the cross section
of the disclination turns during a complete loop. This invariant
applies to elastic loops, DNA loops \cite{dna} and other fields, but
in the case of a $-1/2$ nematic disclination, due to its intrinsic
three-fold symmetry, it assumes specific fractional values, similar to
flux discretization in the fractional quantum Hall effect \cite{qhe}.

For the investigation of the self-linking number, we chose a colloidal
dimer consisting of two spherical particles with strong homeotropic
anchoring, confined to a homogeneous planar nematic cell
\cite{zum1}. Depending on the particle size, confinement and initial
conditions, the particles can interact by arranging themselves into
dipolar or quadrupolar structures \cite{musevic}, or they can be bound
by $-1/2$ disclination loops shared between both particles
\cite{zum2}. The homogeneous director field environment energetically
disfavors linking and knotting, which reveals the more basic rewiring
properties of entangled states.

\begin{figure}
\includegraphics[width=\columnwidth]{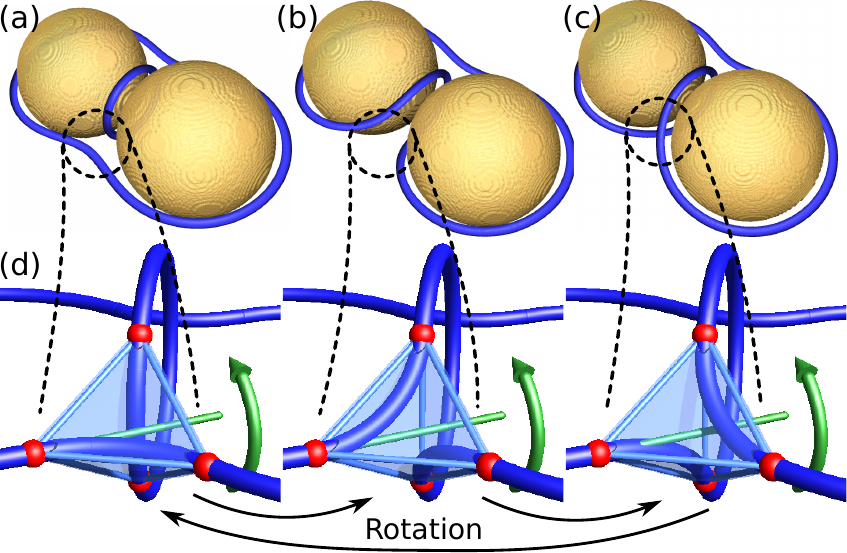}
\caption{\label{fig:figure1}
Rewiring sites of different dimer structures.
{(a-c)} Theta structure and two chiral omega structures (simulations
by M.~Ravnik \cite{zum1}). Rewiring sites are marked and paired with
corresponding idealized structures.
{(d)} The three conformations of a disclination crossing with
tetrahedral symmetry. Rewiring is performed by rotating around a $C_3$
symmetry axis of the tetrahedron.
}
\end{figure}

We demonstrate that differences between dimer structures are localized
to tetrahedral regions around crossings of disclinations
(Fig.~\ref{fig:figure1}), from which we derive rules for calculating
the self-linking number and classifing all dimer structures. The
rewiring rules apply to structures involving $-1/2$ disclinations in
any confinement and can be used to predict and design nematic braids
consisting of complex linked and knotted loops.

We start by examining the similarities of two dimer disclination
configurations \cite{zum2}.  The ``entangled hyperbolic defect
structure'' (referred to as the theta structure from here on) is the
only dimer structure with space inversion symmetry and consists of two
perpendicular loops, one encircling both particles and the other
placed symmetrically between them (Fig.~\ref{fig:figure1}a). The
``omega structure'', on the other hand, consists of a single loop
wrapped around both particles and has two chiral isomers
(Fig.~\ref{fig:figure1}b,c). All three structures have similar
director field and line geometry at the far ends of the colloidal
particles and along the vertical axis (Fig.~5 in \cite{zum2}). They
only differ in the way the left arc, right arc and central loop meet
between the particles (see the encircled areas on
Fig.~\ref{fig:figure1}a-c). The conversion of one structure into
another is achieved by rewiring the crossing, while leaving the remote
field intact, which requires a cutting of disclinations. The resulting
four endpoints define a tetrahedron that encloses the rewiring
site. Inside the tetrahedron, two perpendicular disclination segments
connect pairs of the four vertices
(Fig.~\ref{fig:figure1}d). Experimentally, rewiring is achieved by
local laser melting of a nematic \cite{igor}.

\begin{figure}
\centering
\includegraphics[width=\columnwidth]{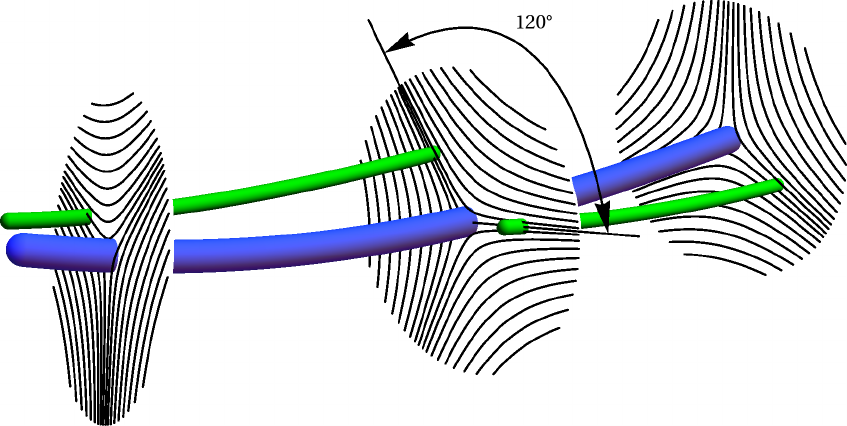}
\caption{\label{fig:figure2}
  To describe a disclination line, we introduce a ribbon, defined by
  an axis curve and a secondary curve that follows the orientation of
  the field cross section of the disclination.  The ribbon may
  reconnect with itself with an offset angle of $\pm 120^\circ$ while
  keeping the director field continuous, due to the symmetry of its
  cross section.
}
\end{figure}

The three-fold symmetry of $-1/2$ disclinations
(Fig.~\ref{fig:figure2}) entering the tetrahedron through the vertices
coincides with the $C_3$ tetrahedral symmetry axes. The director field
inside the tetrahedron has intrinsic dihedral symmetry ($D_{2d}$),
ensured by the relative positioning of the disclinations. Due to this
symmetry and the profile of the disclinations, the director field
stands perpendicularly to all the faces of the tetrahedron and makes
hyperbolic turns at all the edges, thus completing the full
tetrahedral symmetry of the director field on the \emph{surface} of
the tetrahedron. Consequently, rotations from the tetrahedral symmetry
group preserve the continuity of the disclination lines and the
surrounding director field and therefore always generate physically
possible structures. As the disclination segments inside the
tetrahedron have lower symmetry ($D_{2d}$) than the field on its
surface, rotations around a chosen $C_3$ symmetry axis generate 3
distinct configurations of disclinations, depicted in
Fig.~\ref{fig:figure1}d.

The real director field deviates from perfect tetrahedral symmetry in
order to accommodate the proximity of the particles and to minimize
the free energy. However, it only differs from the idealization by a
continuous transformation, so the topological invariants are not
affected. In further derivations, we assume this symmetry to be exact.

The director field surrounding a disclination may rotate around the
disclination line tangent. In a closed loop, rotations are restricted
by the fact that the director field must be continuous. The loop,
together with the orientation of its cross-section, can be described
mathematically by a ribbon (Fig.~\ref{fig:figure2}).
A ribbon can be assigned a self-linking number, $Sl$, a topological
invariant that labels how many times it turns around its tangent in
the course of one loop. Because of the three-fold symmetry of
disclinations, the self-linking number is not restricted to integers,
but can assume any third-integer value (Fig.~\ref{fig:figure2}). Using
of \emph{C\u alug\u areanu theorem}, we can decompose the self-linking
number into writhe and twist \cite{geom1,writh},
\begin{equation}
Sl=Wr+Tw.
\label{eq:calu}
\end{equation}
Writhe depends on how the loop changes direction in space, while twist
contains information about the local torsion of the ribbon around its
axis. Consider the theta structure (Fig.~\ref{fig:figure1}a).  Up to
an arbitrary homotopic transformation, the structure is completely
symmetric and both loops are planar. Both twist and writhe therefore
equal zero, which can be verified using the corresponding Gauss
integral definitions \cite{geom1}.  Tetrahedral rotation of a portion
of the ribbon does not change the twist, as it is defined as an
integral of local twist density, which is preserved by rigid
transformations. As we have shown that the theta structure has zero
twist, the same holds for all entangled dimer structures. It follows
from Eq.~(\ref{eq:calu}) that in this idealization the self-linking
number equals the writhe. The physics of nematic liquid crystals is
hidden in the transformation rules for rewiring and is not involved in
the computation of writhe, which only depends on the disclination loop
geometry.
\begin{figure*}
\centering
\includegraphics[width=\textwidth]{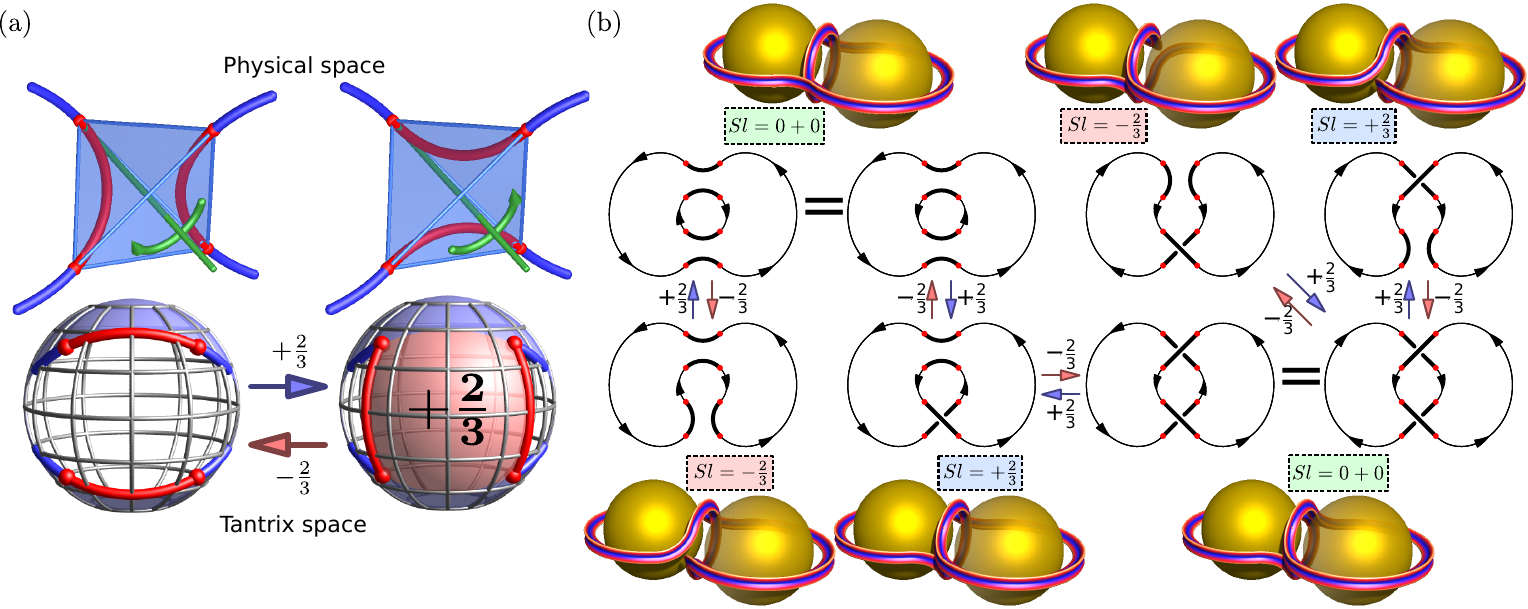}
\caption{\label{fig:figure3}
  {(a)} Tetrahedral rotation changes the positions of two disclination
  line segments, which also changes the curve, traced by the tangent
  on the unit sphere. The spherical area traced by this curve changes
  by $\pm 4\pi/6$ (shown in red), which is directly related to the
  $\pm2/3$ change in writhe and consequently, the self-linking number.
  {(b)} Schematic depiction of dimer structures and transformations
  between them. A tetrahedral rotation changes the writhe by $\pm2/3$
  if parametrizations of initial and final structure are the same. By
  varying the parametrization, we can calculate writhes of all
  different dimer structures. Depicted here are the theta,
  chiral omega and figure-eight structures and the pair of Saturn
  rings, with their respective self-linking numbers. Structures that
  consist of two loops are shown with both possible choices of
  parametrization. Orientations of loop parametrizations are indicated
  by arrows.
}
\end{figure*}
In practice, disclination loops do not necessarily have zero twist,
but this can always be changed by continuous transformations that
preserve the topology of the structure. The twist and writhe convert
into each other under such transformations, but their sum remains
equal to the self-linking number, calculated in our idealized case.

Since the writhe only depends on the axis curve of a ribbon, ordinary
loops can be used instead of ribbons. We use the tantrix
representation: the loop is mapped to the unit sphere of tangents.
Fuller's formula \cite{full} expresses writhe in terms of the
spherical area $A$ enclosed by the tangent indicatrix (tantrix)
loop. The writhe given by this formula has modulo $2$ ambiguity
because full $4\pi$ wraps do not change the tantrix loop,
\begin{equation}
Wr=\frac{A}{2\pi}-1\mod 2.
\label{eq:wrarea}
\end{equation}
The tetrahedral rotations change which pairs of the vertices are
connected by segments of the disclination loop. Each loop segment is
planar and maps to a great circle arc on the tantrix sphere
(Fig.~\ref{fig:figure3}a). The endpoints of these arcs are the
tangents through the vertices of the tetrahedron, which form a square
on the tantrix sphere. Rewiring induced by the tetrahedral rotation
changes the enclosed area by $\mp 4\pi/6$, which by Fuller's formula
(\ref{eq:wrarea}) corresponds to a $\pm2/3$ change in writhe. This is
consistent with the restriction of the self-linking number to thirds,
imposed by the symmetry of $-1/2$ disclinations
(Fig.~\ref{fig:figure2}). It can be shown that the modulo $2$
ambiguity in Eq.~(\ref{eq:wrarea}) can be dropped in our case (see
appendix). As the tangents flip sign if the parametrization of a curve
is reversed, the $\pm2/3$ change in writhe is correct only for
tetrahedral rotations that preserve continuous parametrization of the
loops. If this is not the case, the change in writhe can be calculated
by finding a succession of multiple rewirings that result in the same
structure (Fig.~\ref{fig:figure3}b).

We can generalize the notion of writhe for a union of two or more
loops $A_i$ that may or may not be linked. The Gauss integral that
defines writhe decomposes into writhes of individual loops, which
equal self-linking numbers, $Sl(A_i)$, and linking numbers
$Lk(A_i,A_j)$ between pairs of loops (see appendix)
\begin{equation}
Wr(A_0\cup\cdots\cup A_n)=\sum_{i} Sl(A_i)+2\sum_{i>j}Lk(A_i,A_j).
\label{eq:wrunion}
\end{equation}
Combining this with the fact that every parametrization-preserving
rewiring changes the total writhe by $\pm 2/3$ and changes the number
of loops, $n$, by one \cite{prasolov}, we can write a conservation law
\begin{equation}
\frac{3}{2}(\sum_{i}^n Sl(A_i)+2\sum_{i>j}^n Lk(A_i,A_j))+n=q\mod 2.
\label{eq:law}
\end{equation}
Linking numbers are integers with ambiguously defined sign and the
number of loops may either increase or decrease by one, hence the
modulo $2$. This relation is a generalized conservation of topological
charge $q$\cite{mermin} and reflects the fact that, due to the
presence of line defects, only the even/odd parity of $q$ is conserved
\cite{mermin,janich}. The interpretation of $q$ as the topological
charge can be justified with an example. Consider $q$ homeotropic
spherical particles in a planar nematic cell, each with its own Saturn
ring loop \cite{musevic}. Such a system satisfies the above equation
as it contains $n=q$ unlinked loops with $Sl=0$. Entangled structures
can be reached by applying successive tetrahedral rotations, which
preserve both the left side of Eq.~(\ref{eq:law}) and the topological
charge.

The derived formalism can be demonstrated using our dimer structures.
There are two rewiring sites situated symmetrically between the
colloidal particles. The $3\cdot 3=9$ possible structures consist of
one theta structure, two equivalent structures with disjoint Saturn
rings, two equivalent pairs of chiral omega structures and a pair of
chiral figure-eight structures. The theta structure has zero
self-linking number and the others are reached by successive
tetrahedral rotations. The structures with one loop have self-linking
number $\pm2/3$ while the structures with two loops have both
self-linking numbers equal to zero, which agrees with the conservation
law (\ref{eq:law}). The results are shown in Fig.~\ref{fig:figure3}b.

The derived conservation law (\ref{eq:law}) holds for a set of
multiple linked loops as a whole. Individual constituent loops have a
self-linking number of the form $p/3$, where $p$ is even if an even
number of disclinations pass through the loop and odd in the converse
case. This can be shown by choosing an idealized model of a director
field representing such loop and determining mathematically which
element of the fundamental group the given director field represents
\cite{mermin,janich}. The calculation is carried out in appendix.

We have shown that any rewiring of two orthogonally crossing $-1/2$
disclinations is possible, as the topological requirements are
satisfied entirely by the changes in the self-linking and linking
numbers caused by the application of the tetrahedral rotations. In
contrast, $+1/2$ disclinations cannot form rich entangled structures,
as they only allow integer self-linking numbers. Under the restriction
that only $-1/2$ disclinations are present, the self-linking numbers
of the loops are topological invariants, coupled with surrounding
topological charges by a conservation law. In confined, chiral and
field-affected environments, however, the type of the disclination
profile may vary between $\pm 1/2$ and twist disclinations
\cite{wedge}. Our findings do not apply directly to such cases, as the
self-linking number is ill-defined if the disclination cross-section
does not have constant symmetry.

Our work introduces two important advances in the theoretical
understanding of nematic braids. By combining the formalism of
differential geometry with the characteristics of nematic defects, we
are able to show that the self-linking number is a topological
invariant of $-1/2$ disclination loops that successfully
differentiates between the loops and ensures the conservation of
topological charge. On the other hand, our explanation of local
rewiring by tetrahedral rotations gives a qualitative
three-dimensional image of disclinations and resolves the behavior of
director in complex disclination loop networks seen in experiments and
simulations \cite{araki,zum3,igor}. The richness of entangled
structures increases with the number of available rewiring sites,
which are more abundant in chiral systems \cite{cholesteric,igor}. The
rewiring rules classify the set of possible braids and allow a
transparent design of new structures and guidance of their
experimental realization. Provided symmetry-driven rewiring rules
similar to our tetrahedral rotations exist, our formalism can be
extended to systems with different line defects, crossing geometries
\cite{kang}, or any system with a well-defined self-linking number
(e.g. loop DNA \cite{dna,dnarings}).

We thank T.~Lubensky and R.~Kamien for helpful discussions regarding
the topological interpretation of the results. We acknowledge support
from the Slovenian Research Agency (research program P1-0099 and
project J1-2335) and the NAMASTE Centre of Excellence.

\section{Appendix}

\appendix

\newcommand{\dd}{\,\mathrm{d}}
\newcommand{\ddd}{\mathrm{d}}
\newcommand{\half}{\frac{1}{2}}
\newcommand{\thalf}{\tfrac{1}{2}}
\newcommand{\Tr}{\mathop{\rm Tr}}
\newcommand{\pd}{\partial}
\newcommand{\Dd}[3][{}]{\frac{\ddd^{#1} #2}{\ddd #3^{#1}}}
\newcommand{\Pd}[3][{}]{\frac{\pd^{#1} #2}{\pd #3^{#1}}}
\newcommand{\avg}[1]{\left\langle#1\right\rangle}
\newcommand{\sign}{\mathop{\rm sgn}}
\newcommand{\const}{{\rm const}}
\renewcommand{\vec}[1]{\mathbf{#1}}

In the following sections, we supplement the main article with a
review of the basic knowledge of differential geometry of loops and
ribbons needed to describe general structures consisting of nematic
disclinations. All curves are assumed to be regularly parametrized, so
that their tangents are uniquely defined.

\newcommand{\gaus}{\mathcal{G}}
\section{Gauss map}

\begin{figure*}
  \includegraphics[width=\textwidth]{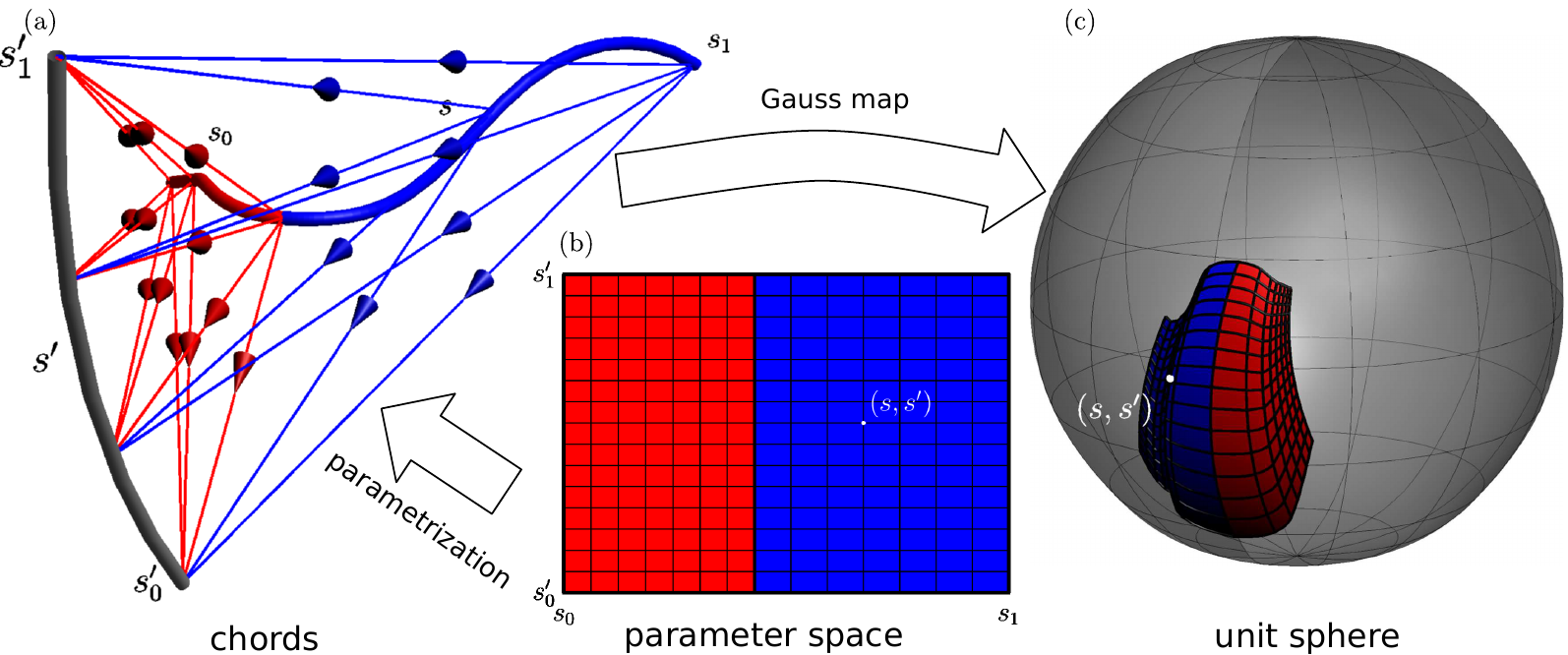}  
  \caption{\label{fig:fig1}
    (a) Two curve segments and a few chords spanning between them.
    Their directions are marked with arrows.
    (b) For two curves in space, a chord between two points is
    defined for every pair of parameters $(s,s')$. The chord directions
    continuously vary with the parameters.
    (c) Directions of chords lie on a patch on the unit
    sphere. Gauss integral calculates the area of this patch. One
    curve is represented as a union of two segments (red and blue),
    which manifests as segmentation of the patch on the unit sphere.
    Gauss integral is bilinear because areas of these patches are
    additive.
  }
\end{figure*}

Consider two nonintersecting curve segments $A$ and $B$ with
parameters $s$ and $s'$. Each pair of points from the two segments
define a chord. For a given parametrization, we can define a Gauss map
that assigns a direction of the chord to each pair of points
(Fig.~\ref{fig:fig1}a). This maps from the set of chords (chord
manifold) to the unit sphere \cite{dennis,geom1}
\begin{equation}
\quad (s,s')\mapsto \frac{\vec{r}_A(s)-\vec{r}_B(s')}{|\vec{r}_A(s)-\vec{r}_B(s')|}\in S^2
\quad .
\label{eq:chord}
\end{equation}
A rectangle in parameter space is mapped to a patch on the unit sphere
(Fig.~\ref{fig:fig1}). The area of the patch is obtained by
integrating the Jacobian of the Gauss map, which yields a Gauss
integral.
\begin{equation}
\gaus(A,B)=\frac{1}{4\pi}\iint  \vec{t}_A(s)\times\vec{t}_B(s')\cdot \frac{\vec{r}_A(s)-\vec{r}_B(s')}{|\vec{r}_A(s)-\vec{r}_B(s')|^3}\dd s\dd s'
\label{eq:gdef}
\end{equation}
From above definition, it is evident that the Gauss integral is
commutative and bilinear under union of curve segments
(Fig.~\ref{fig:fig1}c):
\begin{equation}
\gaus(A,B\cup C)=\gaus(A,B)+\gaus(A,C)\quad .
\label{eq:bilinear}
\end{equation}
If the curves $A$ and $B$ are closed (we will refer to closed curves
as loops), the parameters are cyclic and therefore the chord manifold
has no boundary (it has topology of a torus). The patch on the unit
sphere also has no boundary, so its area can only be an integer
multiple of $4\pi$, depending on how many times the map wraps the
sphere. The Gauss integral of two loops defines the linking number
$Lk(A,B)=\gaus(A,B)$, which is the number of times the loops pass
through each other \cite{geom1,fuller,writh}.

Instead of a simple loop, consider a ribbon, an infinitesimally narrow
two-dimensional strip, seamlessly closed into a loop. A ribbon can be
specified by its two boundary loops: an axis loop and a secondary loop
that differs from the axis loop by a infinitesimal displacement
\cite{dennis}. A ribbon can twist around its tangent, which is
suitable for description of $-1/2$ nematic disclination lines. The
linking number of the axis loop and the secondary loop labels the
ribbons according to the number of twists incorporated in the loop and
is invariant under continuous transformations. To differentiate the
linking number of a ribbon from the linking number of two arbitrary
curves, we will label it with symbol $Sl$ and call it a self-linking
number (this term is used in the field of theoretical geometry in a
narrower context \cite{white}).  Even though both axis loop and
secondary loop of the ribbon have to be continous for the self-linking
number to be defined, we can assign fractional self-linking numbers to
a disclination loop, if the ribbon representing the disclination line
runs around the disclination loop multiple times. In the case of
nematics, the three-fold symmetry of $-1/2$ disclinations restricts
the linking number to third-integer values.

The C\u{a}lug\u{a}reanu theorem (\ref{eq:calu}) decomposes the
Self-linking number into twist that can be seen as contribution of
locally distributed torsion and writhe that measures nonplanarity of
the curve.  The writhe is a Gauss integral of the ribbon's axis loop
with itself, $Wr(A)=\gaus(A,A)$.  The chord manifold in this case is
not a torus but an annulus, which has two boundaries, so unlike the
linking number, the writhe can assume any value \cite{dennis}. The
twist, on the other hand, is defined as a single integral along the
loop and the integrand can be interpreted as a local twist
density\cite{geom1}.
\begin{equation}
Tw=\frac{1}{2\pi}\oint  \vec{t}(s)\cdot (\vec{u}(s)\times \partial_s\vec{u}(s))\dd s
\label{eq:twdef}
\end{equation}
$\vec{u}$ is a normalized vector, perpendicular to the axis loop, that
defines the orientation of the ribbon's cross section \cite{geom1}.
Writhe and twist are not topological invariants, but the self-linking
number is, which means that homotopic transformations of the ribbon
only convert between writhe and twist, preserving their sum.

\section{Writhe and the tetrahedral rotations}
We have shown that the tetrahedral rotation changes the writhe by
$\pm2/3$. We used Fuller's formula (\ref{eq:wrarea}), that does not
take into account the entire Gauss integral, but only maps the
boundaries of the chord manifold. This technique misses full $4\pi$
wraps of the sphere, so the writhe is undetermined up to an integer
multiple of $2$.  With a reasoning described below, we show that the
change of writhe by the tetrahedral rotations is exactly $\pm2/3$,
without an undefined additive offset.

\begin{figure}
  \includegraphics[width=\columnwidth]{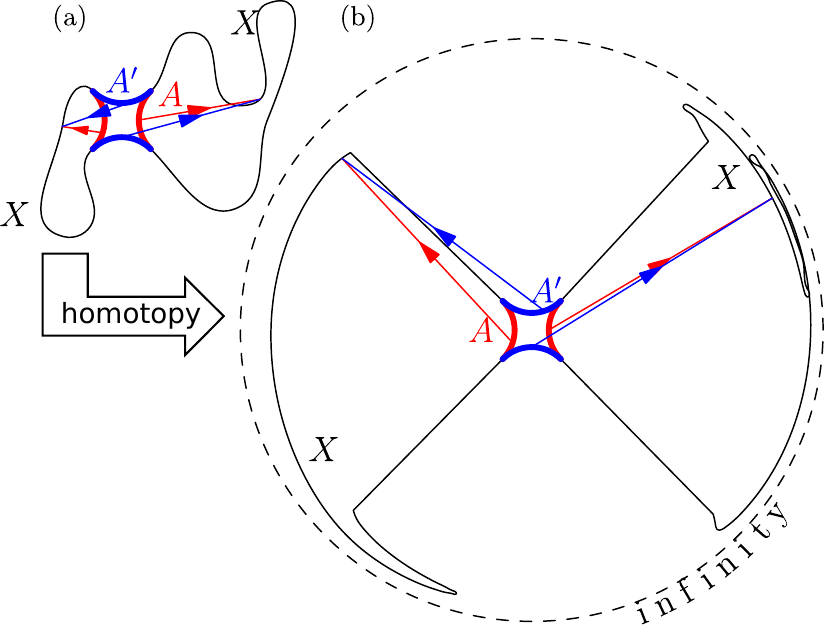}  
  \caption{\label{fig:fig2}
    (a) A tetrahedral rotation transforms segments $A$ into $A'$, which
    only influences the chords between these segments and the rest of the loop $X$.
    (b) We can perform a homotopic transformation that pushes everything but
    the rewiring site to the infinity. In this limit, rewiring for all the curves
    looks exactly the same and the chords connected to the infinity do not
    change direction with rewiring. Note that on this sketch, the chords differ
    because the curve is still at a finite distance from the rewiring site.
  }
\end{figure}

We can represent the loop as a union of two loop segments inside the
tetrahedron $A$ and everything else $X$ (Fig.~\ref{fig:fig2}a). Using
bilinearity (Eq.~\ref{eq:bilinear}) we can expand the result to Gauss
integrals between combinations of segments.
\begin{equation}
Wr(A\cup X)=\gaus(A,A)+\gaus(X,X)+2\gaus(A,X)
\end{equation}
The tetrahedral rotation rigidly rotates the segments $A$ and
preserves $X$, so the only affected term is the last one. This term is
a mapping from a set of chords spanning between $A$ and $X$ to the
space of their directions. In our case, the tetrahedron also contains
nematic director field, which means no other disclinations may be
present inside it. If the rotation is performed continuously, no
disclination lines are crossed by the rotating segments in the
process. We can therefore make a homotopic transformation that
stretches every part of the loop outside the tetrahedron towards the
infinity (Fig~\ref{fig:fig2}b). The chords extending from the loop
segments $A$ to the points at infinity $X$ do not change direction
under finite movements of the loop segments $A$. The only chords that
change are those extending from the loop segments to the straight
paths connecting the tetrahedral vertices to the infinity. These paths
are independent from the rest of the loop ($X$), so the change in
writhe for tetrahedral rotation is universal.  The largest angle for
which the chords can change their direction is $\pi/3$, which happens
for the limiting case at the endpoints of the disclination
segments. Together with the fact that most chords do not change at
all, this is not enough to change the area of the patch on the unit
sphere for $4\pi$. Our result that rewiring changes the writhe for
$\pm2/3$ is therefore correct, without the modulo $2$ ambiguity
suggested by Fuller's formula.

\section{Homotopy of disclination loops with nonzero self-linking number}
To investigate the topological properties of loops with nonzero
self-linking number, we have the liberty to choose any configuration
with desired self-linking number, as homotopic transformations do not
influence the result. We choose a planar ring-shaped disclination loop
with radius $1$ and parametrize the surrounding space in toroidal
coordinates with the main angle $\psi$ and inner angle $\theta$. The
basis vectors of a crossection perpendicular to the disclination line
are $\hat{e}_{\psi}=\cos\psi\hat{e}_x+\sin\psi\hat{e}_y$ and
$\hat{e}_z$.

We are interested in disclination lines with winding number $-1/2$,
which is the number of times the director rotates when we encircle the
loop. We start by choosing director field for the cross section at
$\psi=0$.
\begin{equation}
\vec{n}(\theta,0)=\cos(-\tfrac{\theta}{2}) \hat{e}_{\psi=0}+\sin(-\tfrac{\theta}{2})\hat{e}_z
\end{equation}
To describe a disclination loop with nonzero self-linking number $Sl$,
this cross section must rotate rigidly $Sl$-times when $\psi$
increases to $2\pi$. By evaluating these rotations, we obtain full
specification of the director field around the disclination loop.
\begin{equation}
\vec{n}(\theta,\psi)=\cos(\tfrac{1}{2}(\theta-3Sl\psi))\hat{e}_{\psi}-\sin(\tfrac12(\theta-3Sl\psi))\hat{e}_z
\end{equation}
To describe the topology of the director field on a torus, we need two
winding numbers\cite{janich}. The winding number of the small toroidal
circle we already know: it equals $-1/2$. The winding number on the
great toroidal circle is determined by the mapping $\vec{n}(0,\psi)$,
where we fixed the value of $\theta=0$, and measures the homotopy
class of the director field encircled by our disclination loop.
\begin{equation}
\vec{n}(0,\psi)=(\cos(\tfrac{3}{2}Sl\psi)\cos\psi,\cos(\tfrac{3}{2}Sl\psi)\sin\psi,\sin(\tfrac{3}{2}Sl\psi))
\end{equation}
Fundamental group of nematic order parameter only has two elements:
integer winding numbers correspond to a director field, homotopic to
defect-free field and half-integer, which corresponds to defect
lines. We know that in the first case, the vector representation of
the director field is continuous and in the latter case, it has a sign
discontinuity between $\vec{n}(0,0)$ and $\vec{n}(0,2\pi)$.  Assuming
$Sl=p/3$, we get the condition $\cos(p\pi)=\pm 1$. Self-linking
numbers
$Sl=\{\ldots,-\tfrac{2}{3},0,\tfrac{2}{3},\tfrac{4}{3},\ldots\}$
therefore correspond to director field, homotopic to defect-free
field, which is achieved if even number of disclination lines pass
through our loop. Self-linking numbers
$Sl=\{\ldots,-1,-\tfrac{1}{3},\tfrac{1}{3},1,\ldots\}$ arise in case
odd number of disclination lines pass through our loop. For two loops,
their linking number measures how many times they pass through each
other and therefore chooses between two distinct sets of possible
self-linking numbers described above.

\bibliography{CoparZumer-rewiringBraids}

\begin{thebibliography}{29}%
\makeatletter
\providecommand \@ifxundefined [1]{%
 \@ifx{#1\undefined}
}%
\providecommand \@ifnum [1]{%
 \ifnum #1\expandafter \@firstoftwo
 \else \expandafter \@secondoftwo
 \fi
}%
\providecommand \@ifx [1]{%
 \ifx #1\expandafter \@firstoftwo
 \else \expandafter \@secondoftwo
 \fi
}%
\providecommand \natexlab [1]{#1}%
\providecommand \enquote  [1]{``#1''}%
\providecommand \bibnamefont  [1]{#1}%
\providecommand \bibfnamefont [1]{#1}%
\providecommand \citenamefont [1]{#1}%
\providecommand \href@noop [0]{\@secondoftwo}%
\providecommand \href [0]{\begingroup \@sanitize@url \@href}%
\providecommand \@href[1]{\@@startlink{#1}\@@href}%
\providecommand \@@href[1]{\endgroup#1\@@endlink}%
\providecommand \@sanitize@url [0]{\catcode `\\12\catcode `\$12\catcode
  `\&12\catcode `\#12\catcode `\^12\catcode `\_12\catcode `\%12\relax}%
\providecommand \@@startlink[1]{}%
\providecommand \@@endlink[0]{}%
\providecommand \url  [0]{\begingroup\@sanitize@url \@url }%
\providecommand \@url [1]{\endgroup\@href {#1}{\urlprefix }}%
\providecommand \urlprefix  [0]{URL }%
\providecommand \Eprint [0]{\href }%
\providecommand \doibase [0]{http://dx.doi.org/}%
\providecommand \selectlanguage [0]{\@gobble}%
\providecommand \bibinfo  [0]{\@secondoftwo}%
\providecommand \bibfield  [0]{\@secondoftwo}%
\providecommand \translation [1]{[#1]}%
\providecommand \BibitemOpen [0]{}%
\providecommand \bibitemStop [0]{}%
\providecommand \bibitemNoStop [0]{.\EOS\space}%
\providecommand \EOS [0]{\spacefactor3000\relax}%
\providecommand \BibitemShut  [1]{\csname bibitem#1\endcsname}%
\let\auto@bib@innerbib\@empty
\bibitem [{\citenamefont {Poulin}\ \emph {et~al.}(1997)\citenamefont {Poulin},
  \citenamefont {Stark}, \citenamefont {Lubensky},\ and\ \citenamefont
  {Weitz}}]{poulin}%
  \BibitemOpen
  \bibfield  {author} {\bibinfo {author} {\bibfnamefont {P.}~\bibnamefont
  {Poulin}}, \bibinfo {author} {\bibfnamefont {H.}~\bibnamefont {Stark}},
  \bibinfo {author} {\bibfnamefont {T.~C.}\ \bibnamefont {Lubensky}}, \ and\
  \bibinfo {author} {\bibfnamefont {D.~A.}\ \bibnamefont {Weitz}},\ }\href@noop
  {} {\bibfield  {journal} {\bibinfo  {journal} {Science}\ }\textbf {\bibinfo
  {volume} {275}},\ \bibinfo {pages} {1770} (\bibinfo {year}
  {1997})}\BibitemShut {NoStop}%
\bibitem [{\citenamefont {Muševič}\ \emph {et~al.}(2006)\citenamefont
  {Muševič}, \citenamefont {Škarabot}, \citenamefont {Tkalec}, \citenamefont
  {Ravnik},\ and\ \citenamefont {Žumer}}]{musevic}%
  \BibitemOpen
  \bibfield  {author} {\bibinfo {author} {\bibfnamefont {I.}~\bibnamefont
  {Muševič}}, \bibinfo {author} {\bibfnamefont {M.}~\bibnamefont
  {Škarabot}}, \bibinfo {author} {\bibfnamefont {U.}~\bibnamefont {Tkalec}},
  \bibinfo {author} {\bibfnamefont {M.}~\bibnamefont {Ravnik}}, \ and\ \bibinfo
  {author} {\bibfnamefont {S.}~\bibnamefont {Žumer}},\ }\href@noop {}
  {\bibfield  {journal} {\bibinfo  {journal} {Science}\ }\textbf {\bibinfo
  {volume} {18}},\ \bibinfo {pages} {954} (\bibinfo {year} {2006})}\BibitemShut
  {NoStop}%
\bibitem [{\citenamefont {Lapointe}\ \emph {et~al.}(2009)\citenamefont
  {Lapointe}, \citenamefont {Mason},\ and\ \citenamefont
  {Smalyukh}}]{lapointe}%
  \BibitemOpen
  \bibfield  {author} {\bibinfo {author} {\bibfnamefont {C.~P.}\ \bibnamefont
  {Lapointe}}, \bibinfo {author} {\bibfnamefont {T.~G.}\ \bibnamefont {Mason}},
  \ and\ \bibinfo {author} {\bibfnamefont {I.~I.}\ \bibnamefont {Smalyukh}},\
  }\href@noop {} {\bibfield  {journal} {\bibinfo  {journal} {Science}\ }\textbf
  {\bibinfo {volume} {326}},\ \bibinfo {pages} {20} (\bibinfo {year}
  {2009})}\BibitemShut {NoStop}%
\bibitem [{\citenamefont {Hijnen}\ \emph {et~al.}(2010)\citenamefont {Hijnen},
  \citenamefont {Wood}, \citenamefont {Wilson},\ and\ \citenamefont
  {Clegg}}]{cholesteric}%
  \BibitemOpen
  \bibfield  {author} {\bibinfo {author} {\bibfnamefont {N.}~\bibnamefont
  {Hijnen}}, \bibinfo {author} {\bibfnamefont {T.~A.}\ \bibnamefont {Wood}},
  \bibinfo {author} {\bibfnamefont {D.}~\bibnamefont {Wilson}}, \ and\ \bibinfo
  {author} {\bibfnamefont {P.}~\bibnamefont {Clegg}},\ }\href@noop {}
  {\bibfield  {journal} {\bibinfo  {journal} {Langmuir}\ }\textbf {\bibinfo
  {volume} {26}},\ \bibinfo {pages} {13502} (\bibinfo {year}
  {2010})}\BibitemShut {NoStop}%
\bibitem [{\citenamefont {Araki}\ \emph {et~al.}(2011)\citenamefont {Araki},
  \citenamefont {Buscaglia}, \citenamefont {Bellini},\ and\ \citenamefont
  {Tanaka}}]{tanaka2}%
  \BibitemOpen
  \bibfield  {author} {\bibinfo {author} {\bibfnamefont {T.}~\bibnamefont
  {Araki}}, \bibinfo {author} {\bibfnamefont {M.}~\bibnamefont {Buscaglia}},
  \bibinfo {author} {\bibfnamefont {T.}~\bibnamefont {Bellini}}, \ and\
  \bibinfo {author} {\bibfnamefont {H.}~\bibnamefont {Tanaka}},\ }\href@noop {}
  {\bibfield  {journal} {\bibinfo  {journal} {Nat.~Mat.}\ }\textbf {\bibinfo
  {volume} {10}},\ \bibinfo {pages} {303} (\bibinfo {year} {2011})}\BibitemShut
  {NoStop}%
\bibitem [{\citenamefont {Joannopoulos}\ \emph {et~al.}(2008)\citenamefont
  {Joannopoulos}, \citenamefont {Johnson}, \citenamefont {Winn},\ and\
  \citenamefont {Meade}}]{phot}%
  \BibitemOpen
  \bibfield  {author} {\bibinfo {author} {\bibfnamefont {J.~D.}\ \bibnamefont
  {Joannopoulos}}, \bibinfo {author} {\bibfnamefont {S.~G.}\ \bibnamefont
  {Johnson}}, \bibinfo {author} {\bibfnamefont {J.~N.}\ \bibnamefont {Winn}}, \
  and\ \bibinfo {author} {\bibfnamefont {R.~D.}\ \bibnamefont {Meade}},\
  }\href@noop {} {\emph {\bibinfo {title} {Photonic Crystals: Molding the Flow
  of Light}}},\ \bibinfo {edition} {2nd}\ ed.\ (\bibinfo  {publisher}
  {Princeton University Press},\ \bibinfo {year} {2008})\BibitemShut {NoStop}%
\bibitem [{\citenamefont {Ravnik}\ \emph {et~al.}(2007)\citenamefont {Ravnik},
  \citenamefont {Škarabot}, \citenamefont {Žumer}, \citenamefont {Tkalec},
  \citenamefont {Poberaj}, \citenamefont {Babič}, \citenamefont {Osterman},\
  and\ \citenamefont {Muševič}}]{zum1}%
  \BibitemOpen
  \bibfield  {author} {\bibinfo {author} {\bibfnamefont {M.}~\bibnamefont
  {Ravnik}}, \bibinfo {author} {\bibfnamefont {M.}~\bibnamefont {Škarabot}},
  \bibinfo {author} {\bibfnamefont {S.}~\bibnamefont {Žumer}}, \bibinfo
  {author} {\bibfnamefont {U.}~\bibnamefont {Tkalec}}, \bibinfo {author}
  {\bibfnamefont {I.}~\bibnamefont {Poberaj}}, \bibinfo {author} {\bibfnamefont
  {D.}~\bibnamefont {Babič}}, \bibinfo {author} {\bibfnamefont
  {N.}~\bibnamefont {Osterman}}, \ and\ \bibinfo {author} {\bibfnamefont
  {I.}~\bibnamefont {Muševič}},\ }\href@noop {} {\bibfield  {journal}
  {\bibinfo  {journal} {Phys.~Rev.~Lett.}\ }\textbf {\bibinfo {volume} {99}},\
  \bibinfo {pages} {247801} (\bibinfo {year} {2007})}\BibitemShut {NoStop}%
\bibitem [{\citenamefont {Ravnik}\ and\ \citenamefont
  {Žumer}(2009{\natexlab{a}})}]{zum3}%
  \BibitemOpen
  \bibfield  {author} {\bibinfo {author} {\bibfnamefont {M.}~\bibnamefont
  {Ravnik}}\ and\ \bibinfo {author} {\bibfnamefont {S.}~\bibnamefont
  {Žumer}},\ }\href@noop {} {\bibfield  {journal} {\bibinfo  {journal} {Soft
  Matter}\ }\textbf {\bibinfo {volume} {5}},\ \bibinfo {pages} {4520} (\bibinfo
  {year} {2009}{\natexlab{a}})}\BibitemShut {NoStop}%
\bibitem [{\citenamefont {Žumer}(2006)}]{keystone}%
  \BibitemOpen
  \bibfield  {author} {\bibinfo {author} {\bibfnamefont {S.}~\bibnamefont
  {Žumer}},\ }\href@noop {} {\enquote {\bibinfo {title} {Modeling of
  constrained nematic order: from defects to colloidal structures},}\ }\bibinfo
  {howpublished} {Plenary talk at 21${}^{\text{st}}$ Int. Liquid Crystal Conf.,
  Keystone, Colorado} (\bibinfo {year} {2006})\BibitemShut {NoStop}%
\bibitem [{\citenamefont {Araki}\ and\ \citenamefont {Tanaka}(2006)}]{araki}%
  \BibitemOpen
  \bibfield  {author} {\bibinfo {author} {\bibfnamefont {T.}~\bibnamefont
  {Araki}}\ and\ \bibinfo {author} {\bibfnamefont {H.}~\bibnamefont {Tanaka}},\
  }\href@noop {} {\bibfield  {journal} {\bibinfo  {journal} {Phys.~Rev.~Lett.}\
  }\textbf {\bibinfo {volume} {97}},\ \bibinfo {pages} {127801} (\bibinfo
  {year} {2006})}\BibitemShut {NoStop}%
\bibitem [{\citenamefont {Ravnik}\ and\ \citenamefont
  {Žumer}(2009{\natexlab{b}})}]{zum2}%
  \BibitemOpen
  \bibfield  {author} {\bibinfo {author} {\bibfnamefont {M.}~\bibnamefont
  {Ravnik}}\ and\ \bibinfo {author} {\bibfnamefont {S.}~\bibnamefont
  {Žumer}},\ }\href@noop {} {\bibfield  {journal} {\bibinfo  {journal} {Soft
  Matter}\ }\textbf {\bibinfo {volume} {5}},\ \bibinfo {pages} {269} (\bibinfo
  {year} {2009}{\natexlab{b}})}\BibitemShut {NoStop}%
\bibitem [{\citenamefont {Mermin}(1979)}]{mermin}%
  \BibitemOpen
  \bibfield  {author} {\bibinfo {author} {\bibfnamefont {N.~D.}\ \bibnamefont
  {Mermin}},\ }\href@noop {} {\bibfield  {journal} {\bibinfo  {journal}
  {Rev.~Mod.~Phys.}\ }\textbf {\bibinfo {volume} {51}},\ \bibinfo {pages} {591}
  (\bibinfo {year} {1979})}\BibitemShut {NoStop}%
\bibitem [{\citenamefont {Tkalec}\ \emph {et~al.}()\citenamefont {Tkalec},
  \citenamefont {Ravnik}, \citenamefont {Čopar}, \citenamefont {Žumer},\ and\
  \citenamefont {Muševič}}]{igor}%
  \BibitemOpen
  \bibfield  {author} {\bibinfo {author} {\bibfnamefont {U.}~\bibnamefont
  {Tkalec}}, \bibinfo {author} {\bibfnamefont {M.}~\bibnamefont {Ravnik}},
  \bibinfo {author} {\bibfnamefont {S.}~\bibnamefont {Čopar}}, \bibinfo
  {author} {\bibfnamefont {S.}~\bibnamefont {Žumer}}, \ and\ \bibinfo {author}
  {\bibfnamefont {I.}~\bibnamefont {Muševič}},\ }\href@noop {} {\enquote
  {\bibinfo {title} {Knots and links in chiral nematic colloids reconfigurable
  by light},}\ }\bibinfo {note} {To be published}\BibitemShut {NoStop}%
\bibitem [{\citenamefont {Kutter}(2002)}]{polymers}%
  \BibitemOpen
  \bibfield  {author} {\bibinfo {author} {\bibfnamefont {S.}~\bibnamefont
  {Kutter}},\ }\emph {\bibinfo {title} {Elasticity of polymers with internal
  topological constraints}},\ \href@noop {} {Ph.D. thesis},\ \bibinfo  {school}
  {University of Cambridge} (\bibinfo {year} {2002})\BibitemShut {NoStop}%
\bibitem [{\citenamefont {Kamien}(1998)}]{kamwrith}%
  \BibitemOpen
  \bibfield  {author} {\bibinfo {author} {\bibfnamefont {R.~D.}\ \bibnamefont
  {Kamien}},\ }\href@noop {} {\bibfield  {journal} {\bibinfo  {journal}
  {Eur.~Phys.~J.~B}\ }\textbf {\bibinfo {volume} {1}},\ \bibinfo {pages} {1}
  (\bibinfo {year} {1998})}\BibitemShut {NoStop}%
\bibitem [{\citenamefont {Fuller}(1978)}]{full}%
  \BibitemOpen
  \bibfield  {author} {\bibinfo {author} {\bibfnamefont {F.~B.}\ \bibnamefont
  {Fuller}},\ }\href@noop {} {\bibfield  {journal} {\bibinfo  {journal}
  {Proc.~Nat.~Acad.~Sci.~U.S.A.}\ }\textbf {\bibinfo {volume} {75}},\ \bibinfo
  {pages} {3557} (\bibinfo {year} {1978})}\BibitemShut {NoStop}%
\bibitem [{\citenamefont {Han}\ \emph {et~al.}(2010)\citenamefont {Han},
  \citenamefont {Pal}, \citenamefont {Liu},\ and\ \citenamefont {Yan}}]{dna}%
  \BibitemOpen
  \bibfield  {author} {\bibinfo {author} {\bibfnamefont {D.}~\bibnamefont
  {Han}}, \bibinfo {author} {\bibfnamefont {S.}~\bibnamefont {Pal}}, \bibinfo
  {author} {\bibfnamefont {Y.}~\bibnamefont {Liu}}, \ and\ \bibinfo {author}
  {\bibfnamefont {H.}~\bibnamefont {Yan}},\ }\href@noop {} {\bibfield
  {journal} {\bibinfo  {journal} {Nature Nanotech.}\ }\textbf {\bibinfo
  {volume} {5}},\ \bibinfo {pages} {712} (\bibinfo {year} {2010})}\BibitemShut
  {NoStop}%
\bibitem [{\citenamefont {Irvine}\ and\ \citenamefont
  {Bouwmeester}(2008)}]{light}%
  \BibitemOpen
  \bibfield  {author} {\bibinfo {author} {\bibfnamefont {W.~T.~M.}\
  \bibnamefont {Irvine}}\ and\ \bibinfo {author} {\bibfnamefont
  {D.}~\bibnamefont {Bouwmeester}},\ }\href@noop {} {\bibfield  {journal}
  {\bibinfo  {journal} {Nat.~Phys.}\ }\textbf {\bibinfo {volume} {4}},\
  \bibinfo {pages} {716} (\bibinfo {year} {2008})}\BibitemShut {NoStop}%
\bibitem [{\citenamefont {Avron}\ \emph {et~al.}(2003)\citenamefont {Avron},
  \citenamefont {Osadchy},\ and\ \citenamefont {Seiler}}]{qhe}%
  \BibitemOpen
  \bibfield  {author} {\bibinfo {author} {\bibfnamefont {J.~E.}\ \bibnamefont
  {Avron}}, \bibinfo {author} {\bibfnamefont {D.}~\bibnamefont {Osadchy}}, \
  and\ \bibinfo {author} {\bibfnamefont {R.}~\bibnamefont {Seiler}},\
  }\href@noop {} {\bibfield  {journal} {\bibinfo  {journal} {Phys.~Today}\
  }\textbf {\bibinfo {volume} {56}},\ \bibinfo {pages} {38} (\bibinfo {year}
  {2003})}\BibitemShut {NoStop}%
\bibitem [{\citenamefont {Kamien}(2002)}]{geom1}%
  \BibitemOpen
  \bibfield  {author} {\bibinfo {author} {\bibfnamefont {R.~D.}\ \bibnamefont
  {Kamien}},\ }\href@noop {} {\bibfield  {journal} {\bibinfo  {journal}
  {Rev.~Mod.~Phys.}\ }\textbf {\bibinfo {volume} {74}} (\bibinfo {year}
  {2002})}\BibitemShut {NoStop}%
\bibitem [{\citenamefont {Berger}\ and\ \citenamefont {Prior}(2006)}]{writh}%
  \BibitemOpen
  \bibfield  {author} {\bibinfo {author} {\bibfnamefont {M.~A.}\ \bibnamefont
  {Berger}}\ and\ \bibinfo {author} {\bibfnamefont {C.}~\bibnamefont {Prior}},\
  }\href@noop {} {\bibfield  {journal} {\bibinfo  {journal}
  {J.~Phys.~A:~Math.~Gen.}\ }\textbf {\bibinfo {volume} {39}} (\bibinfo {year}
  {2006})}\BibitemShut {NoStop}%
\bibitem [{\citenamefont {Prasolov}\ and\ \citenamefont
  {Sossinsky}(1997)}]{prasolov}%
  \BibitemOpen
  \bibfield  {author} {\bibinfo {author} {\bibfnamefont {V.~V.}\ \bibnamefont
  {Prasolov}}\ and\ \bibinfo {author} {\bibfnamefont {A.~B.}\ \bibnamefont
  {Sossinsky}},\ }\href@noop {} {\emph {\bibinfo {title} {Knots, Links, Braids
  and 3-Manifolds}}}\ (\bibinfo  {publisher} {American Mathematical Society,
  Providence, Rhode Island},\ \bibinfo {year} {1997})\ p.~\bibinfo {pages}
  {32}\BibitemShut {NoStop}%
\bibitem [{\citenamefont {J\"anich}(1987)}]{janich}%
  \BibitemOpen
  \bibfield  {author} {\bibinfo {author} {\bibfnamefont {K.}~\bibnamefont
  {J\"anich}},\ }\href@noop {} {\bibfield  {journal} {\bibinfo  {journal}
  {Acta~Appl.~Math.}\ }\textbf {\bibinfo {volume} {8}},\ \bibinfo {pages} {65}
  (\bibinfo {year} {1987})}\BibitemShut {NoStop}%
\bibitem [{\citenamefont {Fukuda}(2010)}]{wedge}%
  \BibitemOpen
  \bibfield  {author} {\bibinfo {author} {\bibfnamefont {J.}~\bibnamefont
  {Fukuda}},\ }\href@noop {} {\bibfield  {journal} {\bibinfo  {journal}
  {Phys.~Rev.~E}\ }\textbf {\bibinfo {volume} {81}},\ \bibinfo {pages} {040701}
  (\bibinfo {year} {2010})}\BibitemShut {NoStop}%
\bibitem [{\citenamefont {Kang}\ \emph {et~al.}(2001)\citenamefont {Kang},
  \citenamefont {Maclennan}, \citenamefont {Clark}, \citenamefont {Zakhidov},\
  and\ \citenamefont {Baughman}}]{kang}%
  \BibitemOpen
  \bibfield  {author} {\bibinfo {author} {\bibfnamefont {D.}~\bibnamefont
  {Kang}}, \bibinfo {author} {\bibfnamefont {J.~E.}\ \bibnamefont {Maclennan}},
  \bibinfo {author} {\bibfnamefont {N.~A.}\ \bibnamefont {Clark}}, \bibinfo
  {author} {\bibfnamefont {A.~A.}\ \bibnamefont {Zakhidov}}, \ and\ \bibinfo
  {author} {\bibfnamefont {R.~H.}\ \bibnamefont {Baughman}},\ }\href@noop {}
  {\bibfield  {journal} {\bibinfo  {journal} {Phys.~Rev.~Lett.}\ }\textbf
  {\bibinfo {volume} {86}},\ \bibinfo {pages} {4052} (\bibinfo {year}
  {2001})}\BibitemShut {NoStop}%
\bibitem [{\citenamefont {Otto}\ and\ \citenamefont {Vilgis}(1998)}]{dnarings}%
  \BibitemOpen
  \bibfield  {author} {\bibinfo {author} {\bibfnamefont {M.}~\bibnamefont
  {Otto}}\ and\ \bibinfo {author} {\bibfnamefont {T.~A.}\ \bibnamefont
  {Vilgis}},\ }\href@noop {} {\bibfield  {journal} {\bibinfo  {journal}
  {Phys.~Rev.~Lett.}\ }\textbf {\bibinfo {volume} {80}},\ \bibinfo {pages}
  {881} (\bibinfo {year} {1998})}\BibitemShut {NoStop}%
\bibitem [{\citenamefont {Dennis}\ and\ \citenamefont {Hannay}(2005)}]{dennis}%
  \BibitemOpen
  \bibfield  {author} {\bibinfo {author} {\bibfnamefont {M.~R.}\ \bibnamefont
  {Dennis}}\ and\ \bibinfo {author} {\bibfnamefont {J.~H.}\ \bibnamefont
  {Hannay}},\ }\href@noop {} {\bibfield  {journal} {\bibinfo  {journal}
  {Proc.~R.~Soc.~A}\ }\textbf {\bibinfo {volume} {461}},\ \bibinfo {pages}
  {3245} (\bibinfo {year} {2005})}\BibitemShut {NoStop}%
\bibitem [{\citenamefont {Fuller}(1971)}]{fuller}%
  \BibitemOpen
  \bibfield  {author} {\bibinfo {author} {\bibfnamefont {F.~B.}\ \bibnamefont
  {Fuller}},\ }\href@noop {} {\bibfield  {journal} {\bibinfo  {journal}
  {Proc.~Nat.~Acad.~Sci.~U.S.A.}\ }\textbf {\bibinfo {volume} {86}},\ \bibinfo
  {pages} {815} (\bibinfo {year} {1971})}\BibitemShut {NoStop}%
\bibitem [{\citenamefont {White}(1969)}]{white}%
  \BibitemOpen
  \bibfield  {author} {\bibinfo {author} {\bibfnamefont {J.~H.}\ \bibnamefont
  {White}},\ }\href@noop {} {\bibfield  {journal} {\bibinfo  {journal}
  {Am.~J.~Math.}\ }\textbf {\bibinfo {volume} {91}},\ \bibinfo {pages} {693}
  (\bibinfo {year} {1969})}\BibitemShut {NoStop}%
\end{thebibliography}%

\end{document}